\documentclass[twocolumn,showpacs,preprintnumbers,amsmath,amssymb]{revtex4}

\usepackage{graphicx}
\usepackage{dcolumn}
\usepackage{lscape}
\usepackage{bm}
\usepackage{color}

\begin{document}
\preprint{APS/123-QED}

\title{Strong correlation between ferromagnetic superconductivity and pressure-enhanced ferromagnetic fluctuations in UGe$_2$\footnote{Phys. Rev. Lett. {\bf 121}, 237001 (2018).}}

\author{Naoyuki Tateiwa}
\email{tateiwa.naoyuki@jaea.go.jp} 
\author{Yoshinori Haga}%
\author{Etsuji Yamamoto}

\affiliation{
Advanced Science Research Center, Japan Atomic Energy Agency, Tokai, Naka, Ibaraki 319-1195, Japan\\
}
\date{\today}

\begin{abstract}

We have measured magnetization at high pressure in the uranium ferromagnetic superconductor UGe$_2$ and analyzed the magnetic data using Takahashi's spin fluctuation theory. There is a peak in the pressure dependence of the width of the spin fluctuation spectrum in the energy space $T_0$ at $P_x$, the phase boundary of FM1 and FM2 where the superconducting transition temperature $T_{\rm sc}$ is highest. This suggests a clear correlation between the superconductivity and pressure-enhanced magnetic fluctuations developed at $P_x$. The pressure effect on ${T_{\rm Curie}}/{T_0}$, where ${T_{\rm Curie}}$ is the Curie temperature, suggests that the less itinerant ferromagnetic state FM2 is changed to a more itinerant one FM1 across $P_x$. Peculiar features in relations between $T_0$ and $T_{\rm sc}$ in uranium ferromagnetic superconductors UGe$_2$, URhGe and UCoGe are discussed in comparison with those in high-$T_c$ cuprate and heavy fermion superconductors.

 \end{abstract}


\maketitle

Ferromagnetism and usual $s$-wave superconductivity are antagonistic phenomena since the superconducting pairs are easily destroyed by the ferromagnetic exchange field. The coexistence of superconductivity and ferromagnetism both carried by the same electrons has been considered as a fantastic theoretical possibility since
its prediction by Ginzburg\cite{ginzburg}. The $4f$-localized systems such as ErRh$_4$B$_4$\cite{moncton,fertig}, HoMo$_6$S$_8$\cite{ishikawa}, and ErNi$_2$B$_2$C\cite{canfield} show the coexistence of both phases. The ferromagnetism and superconductivity of the systems are carried, however, by different electrons: $f$ and $d$ electrons, respectively, and the phases compete each other. Therefore, the discoveries of the superconductivity in uranium ferromagnets UGe$_2$\cite{saxena,huxley0}, URhGe\cite{aoki0}, and UCoGe\cite{huy} are very interesting since the same $5f$ electrons of the uranium atoms are responsible for both states\cite{huxley1}.

Let us look other systems. Generally, unconventional superconductivity appears around phase boundaries of magnetic phases in strongly correlated electron systems\cite{pfleiderer0,stewart0}. An important issue that could be elucidated experimentally is finding a relation between the magnetism and the superconductivity. Neutron scattering studies have shown relationships between the superconductivity and magnetic excitations in high-$T_{\rm c}$ cuprate\cite{fujita}, iron arsenide\cite{dai}, and heavy fermion superconductors\cite{stockert,aeppli,sato00}. Correlations between the superconductivity and ferromagnetic fluctuations have been studied by nuclear magnetic resonance experiments on UCoGe\cite{thattori1} and URhGe\cite{ytokunaga1}. Compared with the extensive studies of these two compounds, there have been relatively few experimental studies of UGe$_2$ where the superconductivity appears only at high pressure. This is due to the difficulty of making measurements at a very low temperature and high pressure. In this Letter, we report a clear correlation between the superconductivity and pressure-enhanced ferromagnetic fluctuations in UGe$_2$.

    \begin{figure}[b]
\includegraphics[width=7.8cm]{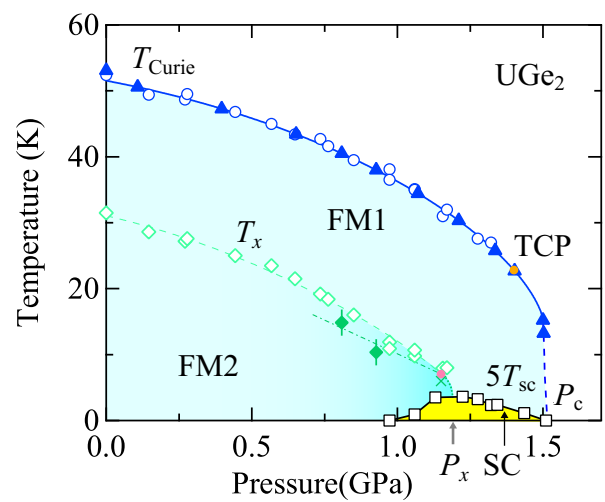}
\caption{\label{fig:epsart}Temperature-pressure phase diagram in UGe$_2$ determined by present magnetic and previous resistivity measurements\cite{tateiwa10}. Dotted and dashed lines for $T_x$ are guides for the eye.}
\end{figure} 

 Figure 1 shows the temperature-pressure phase diagram of UGe$_2$\cite{tateiwa10,tateiwa11}. Open circles and closed triangles represent the Curie temperature $T_{\rm Curie}$ determined by previous resistivity\cite{tateiwa10} and our present magnetic measurements, respectively. $T_{\rm Curie}$ decreases with increasing pressure from 53 K at ambient pressure. The transition changes from the second to first order at a trictical point (TCP: ${T_{\rm TCP}}$ ${\sim}$ 22 K, ${P_{\rm TCP}}$ ${\sim}$ 1.42 GPa), denoted as a filled orange circle, and disappears above a critical pressure ${P_{\rm c}}$ ${\sim}$ 1.5 GPa\cite{huxley1,taufour}. There is an additional boundary ${T_x}$ that splits the ferromagnetic phase into FM2 and FM1. Open diamonds represents ${T_x}$ determined by resistivity measurements\cite{tateiwa10}. The critical pressure of $T_x$ is ${P_x}$ ${\sim}$ 1.20 GPa. The superconductivity appears from approximately 1.0 GPa to ${P_{\rm c}}$. The spontaneous magnetic moment $p_{\rm s}$, the coefficient of the $T^2$-term in the resistivity $A$, and the linear specific heat coefficient $\gamma$ show drastic changes at ${P_x}$\cite{saxena,huxley0,tateiwa10,pfleiderer,tateiwa12}. The difference of Fermi surfaces between FM1 and FM2 was also reported\cite{terashima1,haga}. The microscopic origin of $T_x$ has not been understood yet. The transition between FM1 and FM2 at low temperatures is first order\cite{huxley1}. The first order transition at $T_x$ changes to a crossover at a critical end point (CEP: ${T_{\rm CEP}}$ ${\sim}$ 7 K, ${P_{\rm CEP}}$ ${\sim}$ 1.16 GPa) denoted as a filled magenta circle\cite{huxley1,taufour}. The superconducting transition temperature $T_{\rm sc}$ becomes highest near $P_x$.

 We used a high-quality single crystal of UGe$_2$ with residual resistivity ratio $RRR$ = 600. The details of the sample preparation were reported previously\cite{tateiwa10,tateiwa11}. We have measured magnetization at high pressure with a miniature ceramic-anvil high-pressure cell (MCAC) designed by us for use in a commercial SQUID magnetometer\cite{tateiwa1,tateiwa2,tateiwa3}. We used ceramic anvils with a culet size of 1.8 mm and a Cu-Be gasket with an initial thickness of 0.9 mm. The diameter of the sample space in the gasket was 0.90 mm. A 0.50 $\times$ 0.40 $\times$ 0.50 mm$^3$ single crystal was placed in the sample space with Daphne 7373 as a pressure-transmitting medium\cite{tateiwa4,murata}. The pressure values at low temperatures were determined from the pressure dependence of $T_{\rm sc}$ of Pb placed in the sample space\cite{eiling}.

 The development of longitudinal magnetic fluctuations in UGe$_2$ has been suggested from NMR experiments\cite{kotegawa1,harada1,harada2}. The magnetic data in UGe$_2$ have been analyzed using Takahashi's spin fluctuation theory to study the dynamical magnetic property in FM1 and FM2\cite{takahashi1,takahashi2}. Recently, we have shown the applicability of the theory to most actinide $5f$ electrons ferromagnets\cite{tateiwa21}. We determined the widths of the spin fluctuation spectrum $T_0$ and $T_{\rm A}$ in energy $\omega$ and momentum {\mbox{\boldmath $q$}} spaces, respectively. The mode-mode coupling term $F_1$ was obtained from the slope ${\zeta}$ of the Arrott plot ($M^2$ versus $H/M$ plot) at 2.0 K with the relation $F_1={{N_{\rm A}}^3}(2{{\mu}_{\rm B}})^4/{k_{\rm B}}{\zeta}$, where $N_{\rm A}$ is Avogadoro's number and $k_{\rm B}$ is the Boltzmann constant. $T_0$ and $T_{\rm A}$ can be estimated with the value of $p_{\rm s}$ using Eqs. (1) and (2).

   \begin{eqnarray}
 &&{\left({{T_{\rm C}}\over{T_0}}\right)^{5/6}} = {{p_{\rm s}^2}\over {5{g^2}C_{4/3}}} {\left({15c{F_1}\over{{2}{T_{\rm C}}}}\right)^{1/2}}\\
&&  {\left({{T_{\rm C}}\over{T_{\rm A}}}\right)^{5/3}} = {{p_{\rm s}^2}\over {5{g^2}C_{4/3}}} {\left({{2}{T_{\rm C}}\over{15c{F_1}}}\right)^{1/2}}
 \end{eqnarray}
where $g$ represents the Lande's $g$ factor and $C_{4/3}$ is a constant ($C_{4/3}$ = 1.006089${\,}{\cdot}{\cdot}{\cdot}$)\cite{takahashi1,takahashi2}. 
  \begin{figure}[t]
\includegraphics[width=7.8cm]{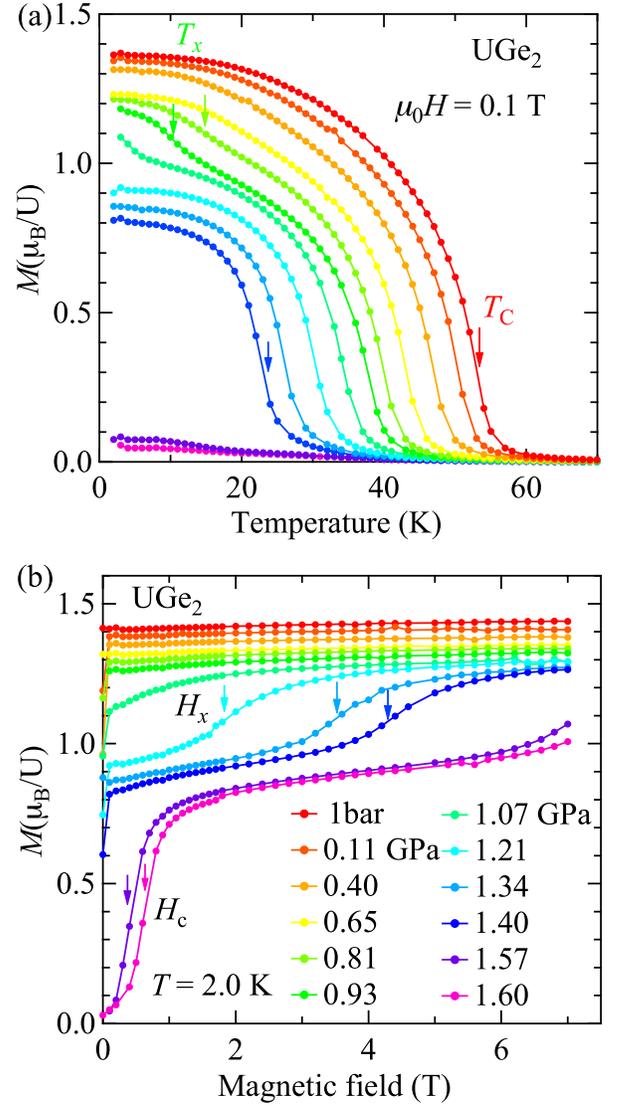}
\caption{\label{fig:epsart}(a)Temperature dependencies of the magnetization in applied magnetic field of 0.1 T and (b) Magnetic field dependencies of the magnetization under ambient and several pressures at $T$ = 2.0 K in UGe$_2$.}
\end{figure} 

The slope ${\zeta}$ of the Arrott plot was determined from the data in a wide magnetic field region up to 7 T since the data points form almost linear straight lines at low temperatures in FM2. Meanwhile, the magnetization $M(H)$ shows the metamagnetic transition at $H_x$ above $P_x$. We analyzed the data up to $H$ = 0.5$-$0.6 $H_x$ below which the linearity of the Arrott plot is fulfilled in FM1.

 Figure 2 (a) shows the temperature dependencies of the magnetization $M(T)$ in a magnetic field of 0.1 T applied along the magnetic easy $a$-axis at ambient and several pressures. $T_{\rm C}$ is defined as the point where $-dM(T)/dT$ is a maximum in a low magnetic field of 0.01 T and is indicated by an arrow in the figure. The pressure dependence of $T_{\rm C}$ is consistent with that determined by the resistivity measurement as shown in Fig. 1\cite{tateiwa10}. The magnetization increases with decreasing temperature monotonically below $T_{\rm C}$ in the low pressure region below 0.65 GPa. A change in the $T$-dependence of $M(T)$ appears at $T_x$ above 0.81 GPa. Closed diamonds in Fig. 1 represents $T_x$ ( = 14.8 and 10.4 K at 0.81 and 0.93 GPa, respectively) defined from the peak position in $-dM(T)/dT$. The value of $T_x$ cannot be determined correctly for 1.07 GPa since the number of the data points at lower temperatures is not enough for the correction determination of $T_x$.  A cross in Fig. 1 represents $T_x$, which is determined in our previous study by the specific heat under high pressure\cite{tateiwa12}.

The values of $T_x$ determined in the present study are slightly lower than those by the resistivity measurement. This difference in $T_x$ might be related to two anomalies in the temperature dependence of the thermal expansion\cite{taufour}. The crossover region of $T_x$ is bound by two lines in the pressure-temperature diagram below $P_{\rm CEP}$. The resistivity shows an anomaly only at the higher temperature line. The plotted data points of $T_x$ in the present study lie close to the lower line\cite{taufour}. The difference in $T_x$ may be an interesting problem, but it is left for the future.  Above $P_x$ where the ground state is FM1, $M(T)$ shows a simple ferromagnetic behavior at 1.21, 1.34 and 1.40 GPa.  The value of the low temperature magnetization becomes less than 1.0 ${{\mu}_{\rm B}}$/U in FM1. $M(T)$ does not show the ferromagnetic behavior at 1.57 and 1.60 GPa, suggesting that the critical pressure $P_{\rm c}$ for the ferromagnetism is about 1.5 GPa.

 Figure 2 (b) shows the magnetic field dependence of the magnetization $M(H)$ at 2.0 K at ambient pressure and several pressures. The magnetization shows a simple ferromagnetic behavior in FM2. The magnetization decreases weakly with increasing pressure below $P_x$.  Above the critical pressure, the value of $p_{\rm s}$ is reduced to less than 1.0 ${{\mu}_{\rm B}}$/U in FM1. $M(H)$ in FM1 increases with increasing magnetic field at low fields and shows an anomalous increase and metamagnetic transition at $H_x$ = 1.80, 3.54, and 4.33 T for 1.21, 1.34, and 1.40 GPa, respectively, where the transition from FM1 to FM2 occurs\cite{tateiwa11}. Above $P_{\rm c}$, the ground state is in the paramagnetic state at zero magnetic field. However, the magnetization increases drastically at $H_{\rm c}$ = 0.42, and 0.65 T for 1.57 and 1.60 GPa, respectively, where FM1 is induced from the paramagnetic state\cite{terashima1,haga}. $M(H)$ increases simply with increasing field and shows a weak nonlinear increase again above 6.0 and 6.4 T at 1.57 and 1.60 GPa, respectively. This suggests the recovery of FM2 for $H{>}H_x$ above 7.0 T.
  
 The decreases of $T_{\rm C}$ and $p_{\rm s}$ under compression suggest a pressure-driven magnetic instability towards the ferromagnetic to paramagnetic quantum phase transition at $P_{c}$\cite{huxley1,brando}. The present results of the magnetic data are basically consistent with those in previous magnetic measurements under high pressure\cite{tateiwa11,pfleiderer,motoyama}. The pressure dependence of $H_x$ is consistent with those in our previous studies\cite{tateiwa11,haga}, but the values of $H_x$ are about 15 \% larger than those in Ref. 23. The reason of the discrepancy is not clear. 
   \begin{figure}[t]
\includegraphics[width=7.8cm]{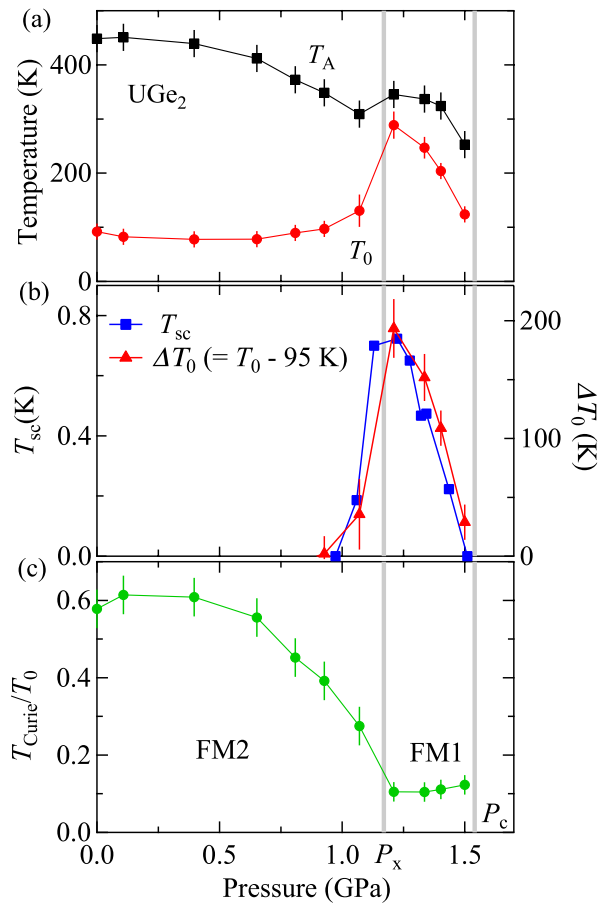}
\caption{\label{fig:epsart}Pressure dependencies of (a) $T_0$ and $T_{\rm A}$, the widths of the spin fluctuations spectrum in energy $\omega$ and momentum {\mbox{\boldmath $q$}} space, respectively, determined from the analysis of the data at 2.0 K, (b) $T_{\rm sc}$(left axis)\cite{tateiwa10} and ${\Delta}{T_0}(P)$ [$={{T_0}(P)}-$95 (K)] (right axis), and (c) ${T_{\rm Curie}}/{T_0}$ in UGe$_2$.}
\end{figure} 

 We analyzed the magnetic data at 2.0 K using Takahashi's spin fluctuation theory. Figure 3 (a) shows the pressure dependencies of spin fluctuation parameters $T_0$ and $T_{\rm A}$: the widths of the spin fluctuations spectrum in energy $\omega$ and momentum {\mbox{\boldmath $q$}} spaces, respectively. $T_0$ and $T_{\rm A}$ show an anomalous enhancement where the superconductivity appears from 1.0 GPa to $P_{\rm c}$. This suggests a change of the spin fluctuation spectrum. There is a clear peak in the pressure dependence of $T_0$ and its peak position is close to $P_x$ where $T_{\rm sc}$ is highest. When the pressure dependence of $T_0$ is expressed as ${{T_0}(P)}={T_0^*}+{\Delta}{T_0}(P)$ where ${T_0^*}$ =  95 K is a pressure-independent term, the pressure dependence of ${\Delta}{T_0}(P)$ scales with that of ${T_{\rm sc}}(P)$ determined by our previous resistivity measurement, as shown in Fig. 3 (b)\cite{tateiwa10}. Theoretical studies have assumed ferromagnetic superconductivity driven by critical fluctuations around a ferromagnetic quantum critical point (QCP)\cite{fay,valls,wang}. This study suggests that the superconductivity in UGe$_2$ is driven by the anomalous magnetic fluctuations with the characteristic energy of 300 K developed around $P_x$. 

We analyzed the magnetic data read from Ref. 23 and determined the pressure dependences of $T_0$ and $T_{\rm A}$. The obtained result is compatible with that in the present paper.

  The drastic changes have been observed in the pressure dependence of $A$, $\gamma$, $p_{\rm s}$ and Fermi surfaces at $P_x$\cite{pfleiderer,tateiwa12,terashima1,haga}. Although several theoretical interpretations have been proposed\cite{sandeman,karchev,watanabe,kubo,wysokinski}, a full microscopic understanding of the transition has remained an open question. Within phenomenological Stoner theory, the magnetic features of the FM1-FM2 transition could be understood if the Fermi surface passes through peaks in the density of states\cite{sandeman}. The pairing interaction ${\lambda}_{\Delta}$ is strongly enhanced at $P_x$ in the Stoner theory\cite{sandeman}. However, the calculated large value of ${\lambda}_{\Delta}$ above $P_c$ in the theory seems not applicable to UGe$_2$ where the superconductivity appears only below $P_c$. Further studies are necessary to elucidate the dynamical magnetic property around $P_x$.

 The ferromagnetism in the uranium ferromagnetic superconductors is carried by the itinerant $5f$ electrons\cite{huxley1,shick1,fujimori}. Here, we discuss differences between FM1 and FM2 from a parameter ${T_{\rm Curie}}/{T_0}$ that reflects the itineracy of the magnetic fluctuations in the spin fluctuation theory\cite{takahashi2}. The smaller value of ${T_{\rm Curie}}/{T_0}$ indicates a weak itinerant ferromagnetism and the local magnetic moment is responsible for the ferromagnetism for ${T_{\rm Curie}}/{T_0}=1$. Figure 3 (c) shows the pressure dependence of ${T_{\rm Curie}}/{T_0}$. ${T_{\rm Curie}}/{T_0}$ is approximately 0.6 below 0.4 GPa in UGe$_2$. The values of ${T_{\rm Curie}}/{T_0}$ and $p_{\rm s}$ (1.41 ${\mu}_{\rm B}$/U at 1 bar) suggest strong itinerant ferromagnetism in FM2. This feature is in contrast with weak itinerant ferromagnetism in URhGe and UCoGe where the values of ${T_{\rm Curie}}/{T_0}$ and $p_{\rm s}$ are 0.121 and 0.41 ${\mu}_{\rm B}$/U, and 0.0065 and 0.039 ${\mu}_{\rm B}$/U, respectively, at 1 bar\cite{tateiwa21,nksato1}. In UGe$_2$, ${T_{\rm Curie}}/{T_0}$ decreases with increasing pressure above 0.6 GPa. The value of the parameter becomes less than 0.3 above 1.0 GPa where the superconductivity starts to appear. ${T_{\rm Curie}}/{T_0}$ shows an almost pressure-independent value of about 0.1 in FM1. This suggests that the less itinerant ferromagnetic state of FM2 in the low pressure region is changed to the more itinerant one of FM1. This pressure dependence of ${T_{\rm Curie}}/{T_0}$ may be related to the changes of the various physical quantities or Fermi surfaces at $P_x$. We suggest that the degree of the itineracy of the $5f$ electrons changes across $P_x$. This result could have relevance to theoretical study with the periodic Anderson model that shows the change of the local $f$ electron state inside the ferromagnetic state\cite{kubo}. It is interesting to note that the value of ${T_{\rm Curie}}/{T_0}$ of FM1 is similar to that in URhGe. A certain degree of itinerancy of the $5f$ electrons might be necessary for the coexistence of the superconductivity and the ferromagnetism.

       \begin{figure}[t]
\includegraphics[width=7.8cm]{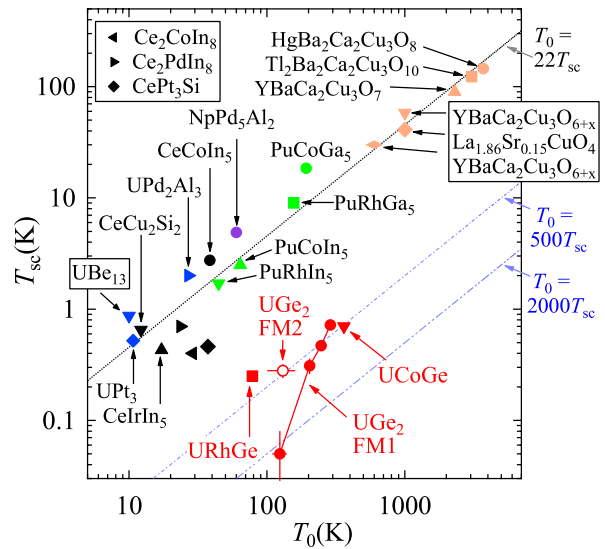}
\caption{\label{fig:epsart}Relations between the superconducting transition temperatures $T_{\rm sc}$ and the energy spread of spin fluctuations $T_0$ for UGe$_2$, URhGe\cite{tateiwa21} and UCoGe\cite{nksato1}, heavy fermion and high-$T_{\rm c}$ cuprate superconductors\cite{moriya7,moriya9,nksato9,nksato2,nksato3}.}
\end{figure}

 Relations between $T_0$ and $T_{\rm sc}$ in UGe$_2$ are discussed quantitatively. $T_{\rm sc}$ is most sensitive to $T_0$ in the strong coupling theory for spin fluctuation-induced superconductivity\cite{moriya8,monthoux1}. The spin fluctuations with higher frequencies are effective for superconductors with high transition temperatures. The correlation between the two quantities has been pointed out in several strongly correlated electron superconductors\cite{moriya7,moriya9,nksato9,nksato2,nksato3}. Figure 4 shows relations between $T_{\rm sc}$ and $T_0$ for UGe$_2$, URhGe\cite{tateiwa21} and UCoGe\cite{nksato1}, heavy fermion superconductors, and high-$T_{\rm c}$ cuprate superconductors. The values of $T_0$ in CePt$_3$Si\cite{takeuchi}, NpPd$_5$Al$_2$\cite{aoki9}, PuRhGa$_5$\cite{bauer9}, PuCoGa$_5$, PuCoIn$_5$, and PuRhIn$_5$ are determined by us from the reported $\gamma$ value with a theoretical expression (${T_0}{\,}{\approx}{\,}{1.25 {\times}{10^4}}/{\gamma}$)\cite{moriya10}. We plot the data of the other systems determined by various experimental methods cited from literatures\cite{moriya7,moriya9,nksato9,nksato2,nksato3}. The data of the cuprate and the heavy fermion superconductors are plotted around a straight dotted line with ${T_0}{\,}={\,}22{T_{\rm sc}}$ denoted as a dotted line in Fig. 4, suggesting a common feature in the superconductivity. $d$-wave superconductivity has been experimentally suggested in a number of superconductors in the strongly correlated electron systems\cite{tsuei,bauer9,pfleiderer2}. Theoretical studies have shown that an optimum frequency ${\omega}_{\rm opt}$ of the antiferromagnetic spin fluctuation spectrum that contributes to raise $T_{\rm sc}$ the most is approximately 10$T_{\rm sc}$ for the $d$-wave superconductivity\cite{monthoux2,mchale}. It is reasonable that the data of the cuprates and heavy fermion superconductors are plotted comparably close to the solid line. Meanwhile, the data for UGe$_2$, URhGe, and UCoGe largely deviate from the relation, suggesting peculiar features in the uranium ferromagnetic superconductors. The data points in FM1 of UGe$_2$ are plotted roughly between lines with ${T_0}{\,}={\,}500{T_{\rm sc}}$ and ${T_0}{\,}= {\,}2000{T_{\rm sc}}$ shown as one and two dot chain lines, respectively. Spin fluctuations with characteristic energy more than two or three orders of magnitude larger than $T_{\rm sc}$ play an important role for the ferromagnetic superconductivity. The values of $T_{\rm sc}$ in FM1 of UGe$_2$ are more than one order of magnitude smaller than those of the $d$-wave superconductors PuCoGa$_5$ and PuRhGa$_5$\cite{bauer9}. Note that the values of $T_0$ in the plutonium superconductors are similar to those in FM1 of UGe$_2$. Theoretical calculation has shown that $T_{\rm sc}$ for $d$-wave pairing in nearly antiferromagnetic metals is about one order magnitude larger than that for the $p$-wave pairing in nearly ferromagnetic metals for comparable conditions such as the band width or strength of the pairing interaction\cite{monthoux3}. Thus, the difference in $T_{\rm sc}$ could be understood if we assume the $p$-wave superconductivity suggested for the uranium ferromagnetic superconductors from anomalous behaviors of the upper critical field $H_{\rm c2}$\cite{khattori,mineev1,wu}. 
 
 The relation between $T_{\rm sc}$ and $T_0$ is expressed as ${T_{\rm sc}}{\,}{\propto}{\,}({T_0})^{\alpha}$ with ${\alpha}$ = 2.3 $\pm$ 0.1 in FM1, which is contrary to the cuprate and heavy fermion superconductor where the linear relation has been discussed. This may reflect unique features in the superconductivity in UGe$_2$. In addition, recent NMR and uniaxial compression studies have suggested the importance of transverse magnetic fluctuations in URhGe\cite{ytokunaga1,braithwaite}. Although the primary parameter that determines $T_{\rm sc}$ is the strength of the longitudinal magnetic fluctuations, it may be necessary to consider the transverse magnetic fluctuations for a complete understanding of the uranium ferromagnetic superconductors\cite{mineev1}.

  In conclusion, we have measured magnetization of UGe$_2$ at high pressure. The analysis of the magnetic data with Takahashi's spin fluctuation theory suggests that the superconductivity in UGe$_2$ is mediated by magnetic fluctuations with characteristic energy of 300 K developing around $P_x$, the first order phase boundary of FM1 and FM2 where $T_{\rm sc}$ is highest. The pressure dependence of ${T_{\rm Curie}}/{T_0}$ suggests that the less itinerant ferromagnetic state FM2 is changed to the more itinerant one FM1 across $P_x$. Peculiar features in the relations between $T_0$ and $T_{\rm sc}$ in uranium ferromagnetic superconductors UGe$_2$, URhGe, and UCoGe are discussed in comparison with those in high-$T_{c}$ cuprate and heavy fermion superconductors.

We thank Prof. Z. Fisk for enlightening suggestions and his editing of this paper. This work was supported by JSPS KAKENHI Grant No. JP16K05463. 

\bibliography{apssamp}

\end{document}